\documentclass{aastex}
\begin{document}
\title{The Interplanetary Network Supplement to the BATSE 5B Catalog of Cosmic
Gamma-Ray Bursts}
\author{K. Hurley}
\affil{University of California, Berkeley, Space Sciences Laboratory,
Berkeley, CA 94720-7450}
\email{khurley@sunspot.ssl.berkeley.edu}
\author{M. S. Briggs}
\affil{University of Alabama in Huntsville, Huntsville AL 35899}
\author{R. M. Kippen}
\affil{Los Alamos National Laboratory, Los Alamos, NM 87545}
\author{C. Kouveliotou\altaffilmark{1}, C. Meegan, G. Fishman} 
\affil{NASA Marshall Space Flight Center, Huntsville AL 35812}
\author{T. Cline}
\affil{NASA Goddard Space Flight Center, Code 661, Greenbelt, MD 20771}
\author{J. Trombka, T. McClanahan}
\affil{NASA Goddard Space Flight Center, Code 691, Greenbelt, MD 20771}
\author{W. Boynton}
\affil{University of Arizona, Lunar and Planetary Laboratory, Tucson, AZ 85721}
\author{R. Starr}
\affil{The Catholic University of America, Department of Physics, Washington, DC 20064}
\author{R. McNutt}
\affil{Johns Hopkins University, Applied Physics Laboratory, Laurel, MD 20723}
\author{M. Boer}
\affil{Centre d'Etude Spatiale des Rayonnements, B.P. 4346, 31029
Toulouse, France}

\altaffiltext{1}{Universities Space Research Association, Marshall Space
Flight Center ES-84, Huntsville, AL 35812}

\begin{abstract}

We present Interplanetary Network (IPN) localization information for 343 gamma-ray bursts
observed by the Burst and Transient Source Experiment (BATSE) between the end of 
the 4th BATSE catalog and the end of the \it Compton Gamma-Ray Observatory \rm (CGRO) mission, obtained by analyzing the arrival times of
these bursts at the \it Ulysses\rm, \it Near Earth Asteroid Rendezvous\rm
(NEAR), and CGRO spacecraft.  
For any given
burst observed by CGRO and one other spacecraft, arrival time analysis
(or ``triangulation'') results in an annulus of possible arrival
directions whose half-width varies between 11 arcseconds and 21 degrees, depending
on the intensity, time history, and arrival direction of the burst, 
as well as the distance between the 
spacecraft.  This annulus generally
intersects the BATSE error circle, resulting in an average reduction of the
area of a factor of 20.  
When all three spacecraft observe a burst, the result is an error
box whose area varies between 1 and 48000 square arcminutes, resulting
in an average reduction of the BATSE error circle area of a factor of 87. 
\end{abstract}


\section{Introduction}

This paper presents the 9th catalog of gamma-ray burst (GRB) localizations 
obtained by arrival time analysis, or ``triangulation'' between the \it
Ulysses \rm spacecraft and other missions in the 3rd interplanetary network (IPN).
Three of these catalogs (Hurley et al. 1999a,b, 2005) were supplements to the BATSE
3B and 4Br burst catalogs (Meegan et al. 1996; Paciesas et al. 1998) and
to the catalogs of untriggered bursts published by Kommers et al. (1999)
and Stern et al. (2000).  The others
involved bursts observed by numerous other spacecraft (Laros et al. 1997, 1998;
Hurley et al., 2000a,b,c).  In
the present paper, we complete the BATSE triggered event catalog supplements with
the data on 343 bursts which occurred between the end of the 4Br catalog
(trigger 5586 on 1996 August 29) and the end of the CGRO mission (corresponding
to trigger 8121 on 2000 May 26).  The BATSE data on these events appear in a
companion paper (Briggs et al. 2006a).  
As none of the
information in the 3B or 4Br supplements has changed, we do not include any of the
detailed data on those bursts in this catalog.

\section{Instrumentation, Search Procedure, Derivation of Annuli, and
Burst Selection Criteria}

With one exception, which we discuss below (the addition of the
NEAR spacecraft to the IPN), these have not changed.  We review each briefly,
but refer the reader to Hurley et al. (1999a,b) for more detailed descriptions.

The \it Ulysses \rm GRB detector (Hurley et al. 1992) consists of two 3 mm thick
hemispherical CsI scintillators
with a projected area of about 20 cm$^2$ in any direction.  The detector is mounted
on a magnetometer boom far from the body of the spacecraft, and has a practically
unobstructed view of the full sky.  During the period covered by
this catalog, the \it Ulysses \rm-Earth distance
varied between about 1500 and 3100 light-seconds.
BATSE consisted of eight detector modules situated at the corners of the \it Compton
Gamma-Ray Observatory \rm spacecraft.  Each contained a Large Area Detector (LAD), 
consisting of a 
50.8 cm diameter by 1.27 cm thick
NaI scintillator  (Meegan et al. 1996).  

The NEAR mission was launched in 1996 February.  Its orbit included an Earth
flyby in 1998 January, and an unsuccessful first attempt to enter into orbit around the
asteroid Eros in 1998 December.  The spacecraft finally went into Eros orbit in
2000 February.  
The mission carried a gamma-ray spectrometer for the study of the surface
composition of Eros (Goldsten et al. 1997; McClanahan et al. 1999; 
Trombka et al. 1999).  Its Bismuth Germanate (BGO) anticoincidence shield is a
right circular cylinder 8.9 cm in diameter by 14 cm long.  Although it was not
specifically designed as a GRB detector, an in-flight modification to the on-board  
software allowed the shield to record the time histories of bursts with 1 s time
resolution in a single energy channel ($\sim$0.15 - 10 MeV), and the first GRB was detected
on 1997 September 2.
During the period covered by this catalog, the NEAR-Earth distance varied between
0.03 and 1300 light-seconds, approximately.

Every cosmic burst detected by BATSE in triggered mode was systematically searched for in the
\it Ulysses \rm and NEAR data by using the approximate arrival direction
from BATSE and the positions of the \it Ulysses \rm and NEAR spacecraft to calculate
a window of possible arrival times at these spacecraft.  Typical window lengths were 300 - 500 s,
and the data were plotted and searched both by eye and software for a count rate
increase corresponding to the BATSE event.

When a GRB arrives at two spacecraft with a delay $\rm \delta$T, it may be
localized to an annulus whose half-angle $\rm \theta$ with respect to the
vector joining the two spacecraft is given by 
\begin{equation}
cos \theta=\frac{c \delta T}{D}
\end{equation}
where c is the speed of light and D is the distance between the two
spacecraft.  (This assumes that the burst is a plane wave, i.e. that its
distance is much greater than D.)  The annulus width d$\rm \theta$, and thus one dimension of
the resulting error box, is 
\begin{equation}
d \rm \theta =c \rm \sigma(\delta T)/Dsin\rm \theta
\end{equation}
where
$\rm \sigma(\delta$T) is the uncertainty in the time delay.  
The radius of each annulus and its right ascension and declination are
calculated in a heliocentric (i.e., aberration-corrected) frame.  These
annuli are given in table 1, and an example is shown in figure 1.  In general, the annuli obtained by triangulations are small circles on the celestial
sphere, so their curvature, even across a relatively small BATSE error circle, is
not always negligible, and a simple, four-sided error box cannot be defined. 
An extreme example is shown in figure 2.
For this reason, we do not cite the intersection points of
the annuli with the error circles.  A prescription for deriving these points,
however, may be found in Hurley et al. (1999a).

When three experiments (\it Ulysses \rm, NEAR, and
BATSE) observe a burst, the result is two annuli which generally 
intersect to define two widely separated error boxes.  The BATSE error circle is used to distinguish the correct
one. An example is shown in figure 3.  

A number of degenerate cases can occur when there is a three-experiment
detection. First, the annuli
can intersect at grazing incidence.  Geometrically, this corresponds to the
case $\rm \theta_1 + \theta_2 \approx \angle NBU $, where $\rm \theta_1$ is the
radius of the first IPN annulus in equation 1, $\rm \theta_2$ is the radius
of the second IPN annulus in equation 1, and $\rm \angle$ NBU is the
NEAR-BATSE-\it Ulysses \rm angle.  When this occurs, the annuli may cross at 
one, two, three, or four points, often defining a long, narrow error box.  An example is
shown in figure 4.  Second,
if the three spacecraft are approximately aligned, 
i.e. $\rm \angle NBU \approx 0 $, one annulus can be completely contained within the other,
i.e. $\rm \theta_1 \approx \theta_2$.  Because NEAR and CGRO are in the ecliptic plane, and \it Ulysses \rm is outside it, this case did not occur for any of the bursts discussed here.  Third,
if the burst arrives along the line joining two spacecraft, $\rm \theta \approx 0 , sin\theta
\approx 0$, and the annulus has a small radius but a very large width $d \rm \theta$.  In
these cases, the width of the annulus can exceed its radius, and the annulus becomes
a circle; the second annulus intersects it to define a single error box.  Finally, if
the annuli intersect almost at grazing incidence, the
two intersections are close to one another and may both be located within the BATSE 1 $\sigma$ error circle,
making it impossible to determine the correct location.  An example is shown in figure 5. 
A total of 11 events produced degenerate localizations.

The main selection criterion for a burst in this catalog is that it 
must have been detected by BATSE and/or \it Ulysses \rm and/or
NEAR.  As in previous catalogs, we also utilized several other criteria, based on the 
goodness-of-fit between the light curves, as judged by the correlation
coefficient and a chi-squared statistic, as well as the
ratio of the observed \it Ulysses \rm counts to the BATSE counts.
There is, however, a difference in the triangulation technique
between the events presented here and those presented in previous catalogs.
GRB time histories are energy-dependent.  A time history taken in the 25-150 keV
\it Ulysses \rm energy range may differ from that taken in the NEAR $\sim$150-10000 keV range.
The magnitude of this difference varies considerably from event to event, and
can easily be judged in, say, the $\chi^2$ technique (Hurley et al. 1999a), where
the goodness-of-fit is reflected in the value of $\chi^2$ per degree of freedom.
When the match between two time histories is poor, the estimate of the statistical uncertainty
in the time difference may become unreliable.  This, in turn, renders the annulus
width estimates, and hence the confidence value for the error box, suspect.
We have been able to avoid this problem here
by comparing time histories in the same, or very similar energy ranges.  Thus
the \it Ulysses \rm 25-150 keV time histories were
compared to 25-100 keV BATSE time histories, while the NEAR time histories were compared to the $>$100 keV BATSE time histories.

\section{A Few Statistics}

There are 1068 bursts in the 5B catalog (Briggs et al. 2006a).  Of these, 
343 were observed by \it Ulysses \rm and/or NEAR, and
in some cases, other spacecraft as well \footnote{.  
A list of all cosmic bursts and the spacecraft which detected them may be found
at http://ssl.berkeley.edu/ipn3/masterli.html or http://heasarc.gsfc.nasa.gov/W3Browse/}.  91 bursts were detected by all
three spacecraft.
Thus the IPN observed
approximately one out of every 3.1 BATSE bursts over this period.  The fraction
of BATSE bursts observed by the IPN is slighly higher in this catalog than
in previous ones due to the addition of the NEAR spacecraft.  The
combination of the 3B and 4Br supplements and the present catalog contains 708 bursts.

The histogram of Figure 6 shows the distribution of annulus half-widths 
(i.e. $\rm \delta R_{IPN}$ in Table 1) for the
343 bursts localized.  The smallest is about 11 \arcsec, the
largest 21$^{\rm o}$, and the average is 9.6 \arcmin.  (The largest annulus
width occurred for BATSE 6580 on 1998 January 25, when the NEAR spacecraft
passed the Earth at a distance of only 0.478 light-seconds on its way to encounter with Eros.)  

Of the 343 bursts, 253 were localized to single annuli by the IPN, and
another 11, although detected by all three spacecraft, were degenerate 
localizations which prevented the derivation of a useful
error box.  These may be treated as single annuli.  Considering these 264 annuli, 185, 
or 70\%, intersect the BATSE 1 $\sigma$ error circles, whose radii are defined by 
$\rm r_{1\sigma}=\sqrt{\sigma_{stat}^2 + \sigma_{sys}^2}$,
where $\sigma_{sys}$ is the systematic error, 1.6$^{\rm o}$,
and $\sigma_{stat}$ is the statistical error.  The combined IPN/BATSE
error regions are an average factor of $\sim$ 20 smaller than the BATSE error circles.
70\% is somewhat less than
the percentage which would be predicted purely on the basis of statistics (87\%).  A
detailed comparison of the IPN and BATSE localizations
results in several more complicated BATSE error
models that render the BATSE and IPN localizations consistent (Briggs et al. 2006b).

Although 91 bursts were observed by \it Ulysses \rm, NEAR, and
BATSE, 11 were degenerate, leaving 80 with well-defined IPN error boxes.  The
smallest has an area of approximately 0.7 square arcminute, and the largest, 48000
square arcminutes.  The average area is 1100 square arcminutes, and
the average ratio of the area of the BATSE error circle to the area of the IPN error box
is $\sim$ 87.  46, or 58\% of these error boxes are partially or completely contained within
their corresponding BATSE error circles.  A histogram of IPN error box sizes is shown in figure 7.
The centers and corners of the 80 error boxes are given in table 2.  However, in some
cases, the curvature of the annuli may render a simple, four-corner error box description
inaccurate.  

Figure 8 shows the BATSE peak fluxes and fluences for 272 of the 343 bursts
with flux and fluence entries in the 5B catalog.  Although the \it Ulysses \rm and NEAR
GRB detectors are roughly 100 and 16 times smaller in area than a single BATSE LAD, figure 8
demonstrates that they are capable of detecting bursts with very small fluences, provided
that the peak flux is relatively high, and vice-versa. 

\section{Tables of Annuli and Error Boxes}

The 14 columns in table 1 give:
1) the date of the burst, in yymmdd format, 
2) the Universal Time of the burst at Earth,
3) the BATSE number for the burst,
4) the BATSE right ascension of the center of the error circle (J2000), in degrees,
5) the BATSE declination of the center of the error circle (J2000), in degrees,
6) the total 1 $\sigma$ statistical BATSE error circle radius, in degrees, (the 
approximate total
1$\sigma$ radius is obtained by adding 1.6$^{\rm o}$ in quadrature, but see
Briggs et al. 2006b for an improved error model), 
7) the right ascension of the center of the first IPN 
annulus, epoch J2000, in the heliocentric frame, in degrees,
8) the declination of the center of the first IPN 
annulus, epoch J2000, in the heliocentric frame, in degrees, 
9) the angular radius of the first IPN annulus, in the heliocentric
frame, in degrees, 
10) the half-width of the first IPN annulus, in degrees; the 3 $\sigma$
confidence annulus is given by R$_{\rm IPN1}$ $\pm$ $\delta$ R$_{IPN1}$,
11) the right ascension of the center of the second IPN 
annulus, epoch J2000, in the heliocentric frame, in degrees,
12) the declination of the center of the second IPN 
annulus, epoch J2000, in the heliocentric frame, in degrees, 
13) the angular radius of the second IPN annulus, in the heliocentric
frame, in degrees, and
14) the half-width of the second IPN annulus, in degrees; the 3 $\sigma$
confidence annulus is given by R$_{\rm IPN2}$ $\pm$ $\delta$ R$_{IPN2}$.

The BATSE data have been taken from the latest online catalog,
and are given here for convenience only;
the catalog \footnote{available at http://gammaray.msfc.nasa.gov/batse/}
should be considered to be the ultimate source of the
most up-to-date BATSE data.   The data in table 1 are also available electronically \footnote
{at http://ssl.berkeley.edu/ipn3/index.html}.  A footnote indicates if the localization
was degenerate, as defined in the previous section, and if the annulus or error box
was confirmed by an independent observation (see the Discussion section).  The
references in the latter footnotes are not meant to be exhaustive; in most cases,
there are numerous reports of follow-up observations.  Most of the references may
be found on the IPN website\footnote
{at http://ssl.berkeley.edu/ipn3/bibliogr.html}.

Table 2 gives the centers and corners of the error boxes for the three-
spacecraft localizations.  The four columns contain:

1) the date of the burst,
2) the BATSE trigger number,
3) the right ascension of the center of the error box, on the first row, and the
right ascensions of the four corners on the following four rows, and
4) the declination of the center of the error box, on the first row, and the 
declinations of the four corners on the following four rows.

All coordinates are J2000.

\section{Discussion and Conclusion}

The years covered in this supplement were witness to a number
of dramatic events which changed the course of GRB studies.  Two missions became
operational which had the capability of independently determining small (several
arcminute) error boxes rapidly, namely the \it Rossi X-Ray Timing Explorer \rm (RXTE), launched
on 1995 December 30, and
\it BeppoSAX \rm, launched on 1996 April 30.  The first GRB counterpart identifications
were made starting in 1997, based on \it BeppoSAX \rm positions, and later identifications
came not only from \it BeppoSAX \rm, but also from RXTE and the IPN.  Apart from the fact that
these identifications were a breakthrough in understanding the origin of gamma-ray
bursts, they were also important because they provided the first localizations which
were as accurate as, or more accurate than the IPN ones.  Thus they serve as a confirmation
of the IPN accuracy.  As noted in table 1, a total of 22 bursts were either localized
independently, confirming the IPN annulus
or error box.  Still another event, the giant flare of 1998 August 27 from SGR1900+14,
and the discovery of its transient radio emission (Hurley et al. 1999c; Frail et al. 1999)
provided a confirmation of the IPN accuracy during this period.

The \it Ulysses \rm mission continues to operate, and the GRB experiment has 
detected over 1800 gamma-ray bursts.  Data on these events may be found at
the IPN web site \footnote{http://ssl.berkeley.edu/ipn3/interpla.html}.  Although
this represents the last installment of the IPN supplements to the BATSE burst catalogs,
work is continuing on the localization of IPN bursts, and on the preparation of catalogs
of these events. 

\section{Acknowledgments}

Support for the \it Ulysses \rm GRB experiment is provided by JPL Contract 958056.  Joint
analysis of \it Ulysses \rm and BATSE data was supported by NASA Grant NAG 5-1560 and NAG5-9701.  NEAR
data analysis was supported under NASA Grants NAG 5-3500 and NAG 5-9503.  We are also
grateful to the NEAR team for their modifications to the XGRS experiment which made
gamma-ray burst detection possible.

\clearpage

\begin{figure}
\plotone{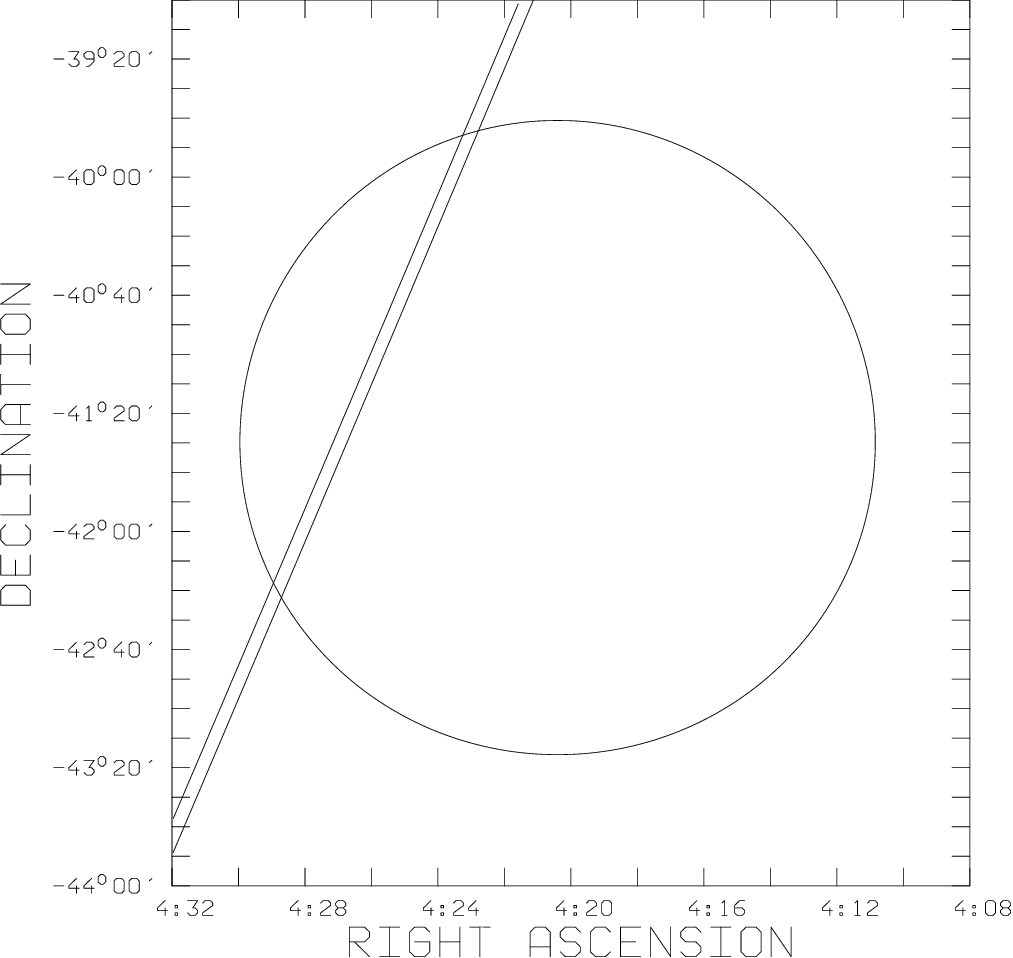}
\caption{The BATSE 1$\sigma$ (statistical + systematic) error circle for
trigger 5593 on 1996 September 6, and the 3$\sigma$ IPN annulus.}
\end{figure}

\begin{figure}
\plotone{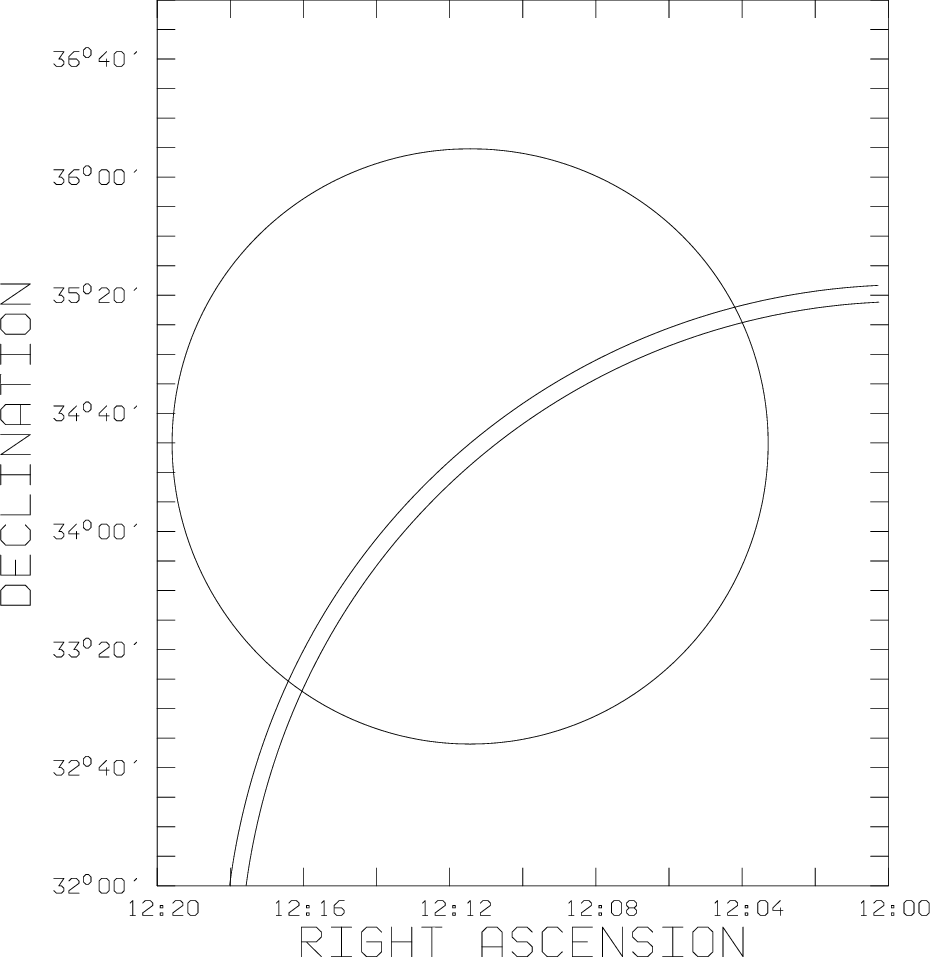}
\caption{The BATSE 1$\sigma$ (statistical + systematic) error circle for
trigger 5711 on 1996 December 12, and the 3$\sigma$ IPN annulus. The
curvature of the annulus makes it impossible to describe the resulting
error box with the four annulus/error circle intersection points.}
\end{figure}

\begin{figure}
\plotone{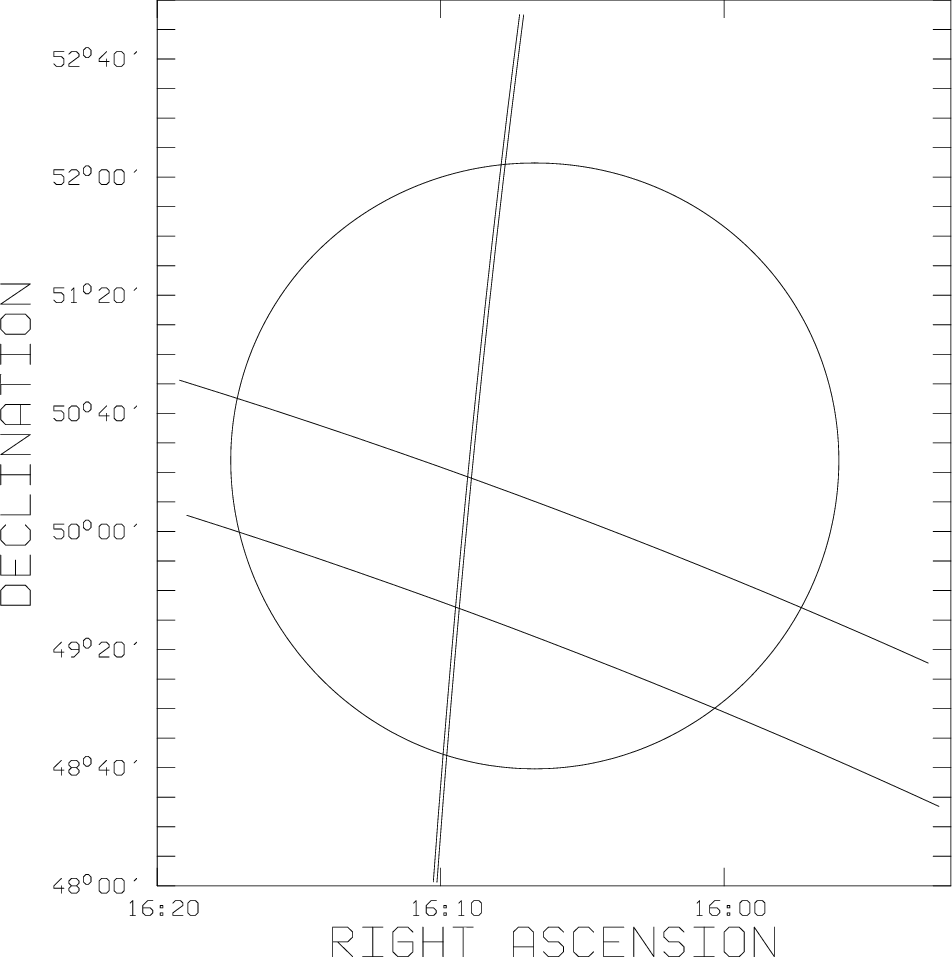}
\caption{The BATSE 1$\sigma$ (statistical + systematic) error circle for
trigger 6472 on 1997 November 10, and the 3$\sigma$ \it Ulysses \rm-BATSE
and NEAR - BATSE annuli. Because NEAR was closer to Earth than Ulysses (smaller
D in equation 2),
and had only 1 s time resolution (resulting in a larger $\rm \sigma(\delta$T) in
equation 2), the NEAR-BATSE annulus is wider.}
\end{figure}

\begin{figure}
\plotone{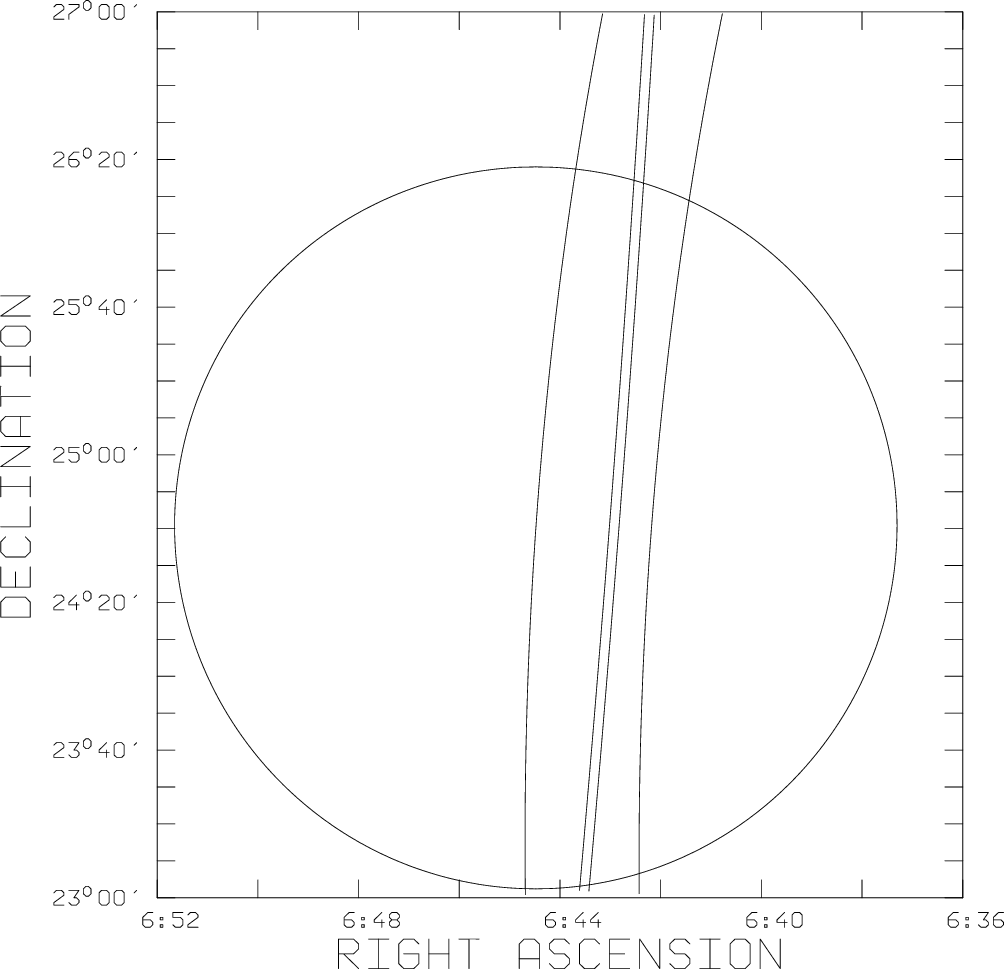}
\caption{The BATSE 1$\sigma$ (statistical + systematic) error circle for
trigger 6453 on 1997 October 29, and the 3$\sigma$ \it Ulysses \rm-BATSE
and NEAR - BATSE annuli. Again, the NEAR-BATSE annulus is wider. Because
the two annuli intersect at grazing incidence, they define a single error
box whose length is much greater than the radius of the BATSE error
circle.}
\end{figure}

\begin{figure}
\plotone{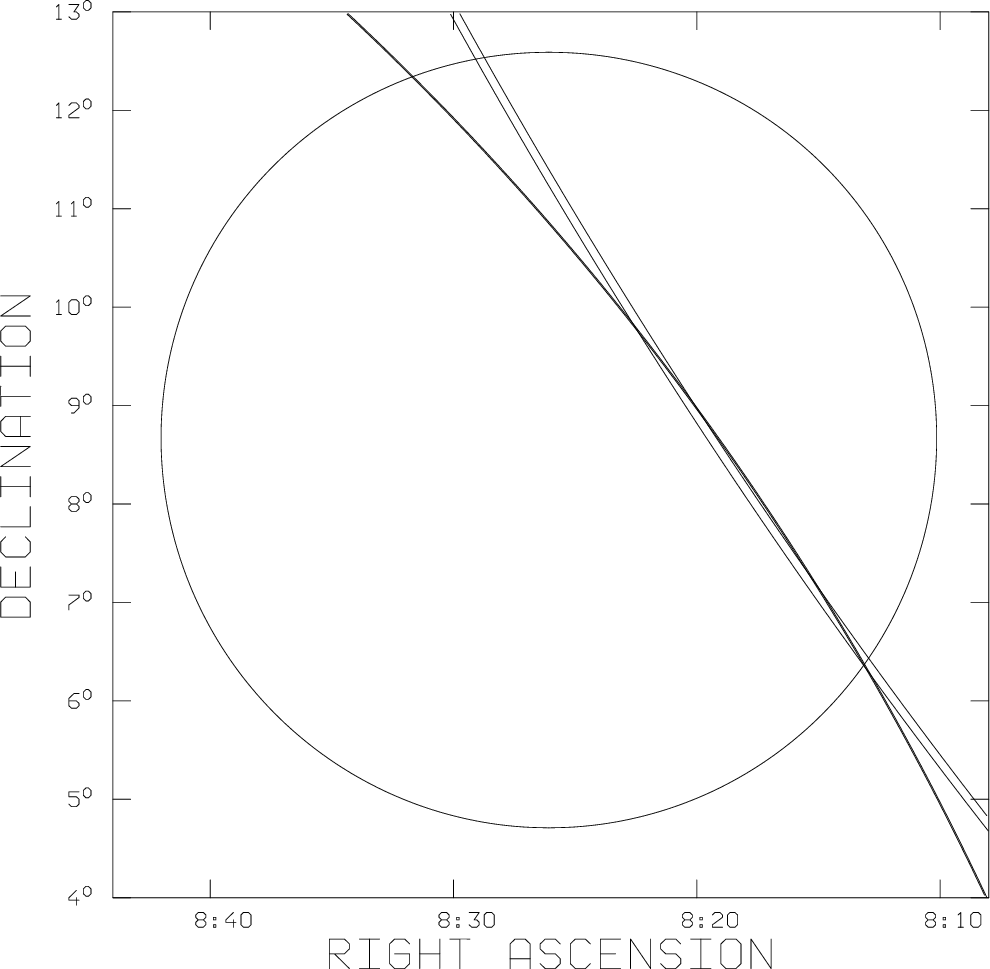}
\caption{The BATSE 1$\sigma$ (statistical + systematic) error circle for
trigger 7647 on 1999 July 12, and the 3$\sigma$ \it Ulysses \rm-BATSE
and NEAR - BATSE annuli. 
the two annuli intersect nearly at grazing incidence, defining two error
boxes within the BATSE error
circle.}
\end{figure}

\begin{figure}
\plotone{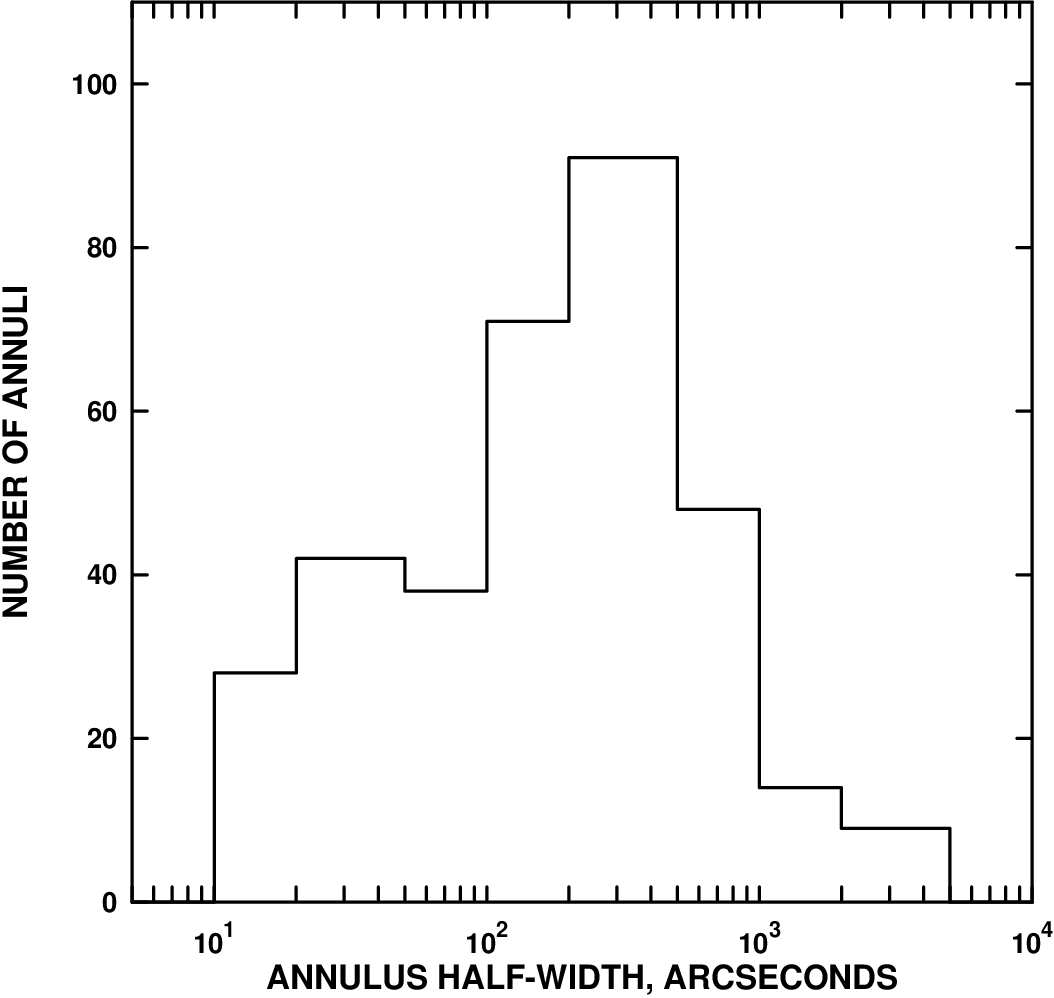}
\caption{Distribution of 343 IPN annulus half-widths. }
\end{figure}

\begin{figure}
\plotone{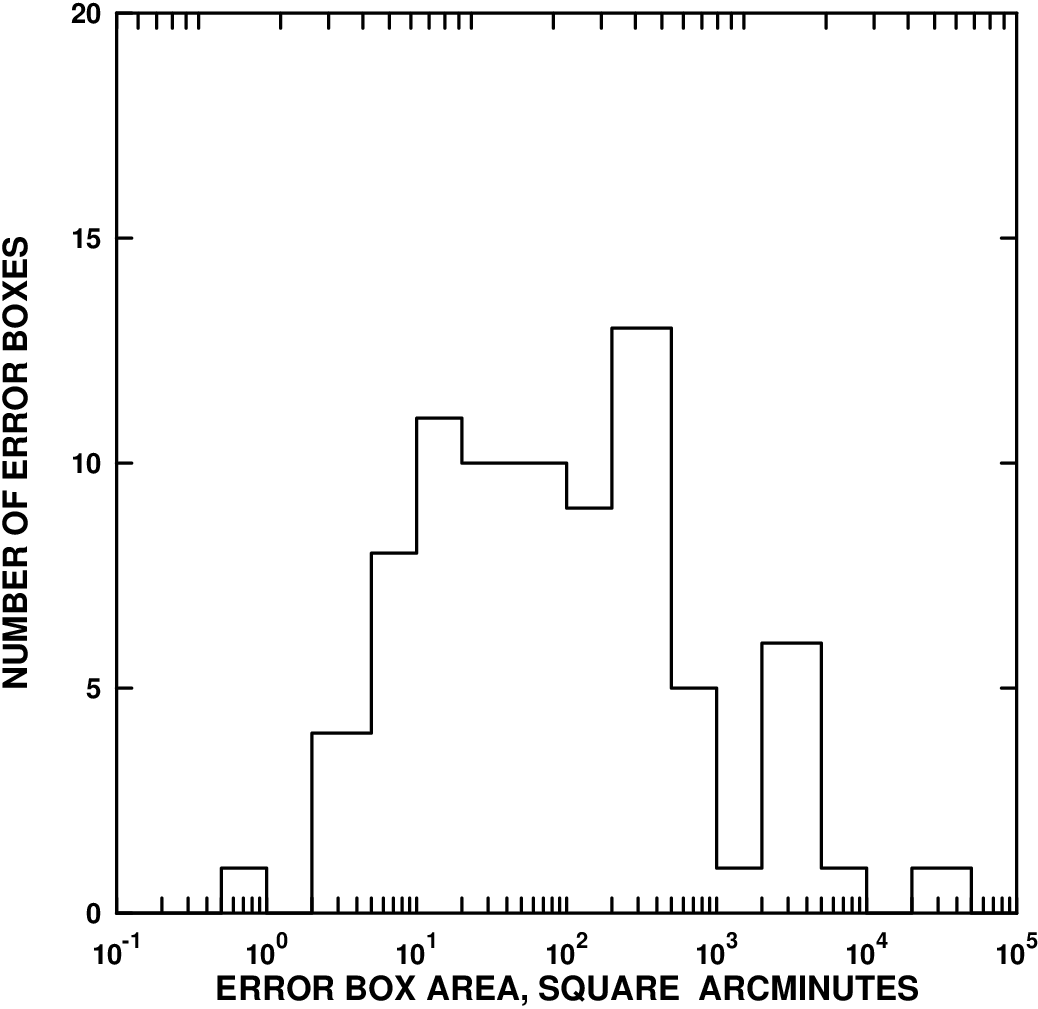}
\caption{Distribution of 80 IPN error box areas.}
\end{figure}

\begin{figure}
\plotone{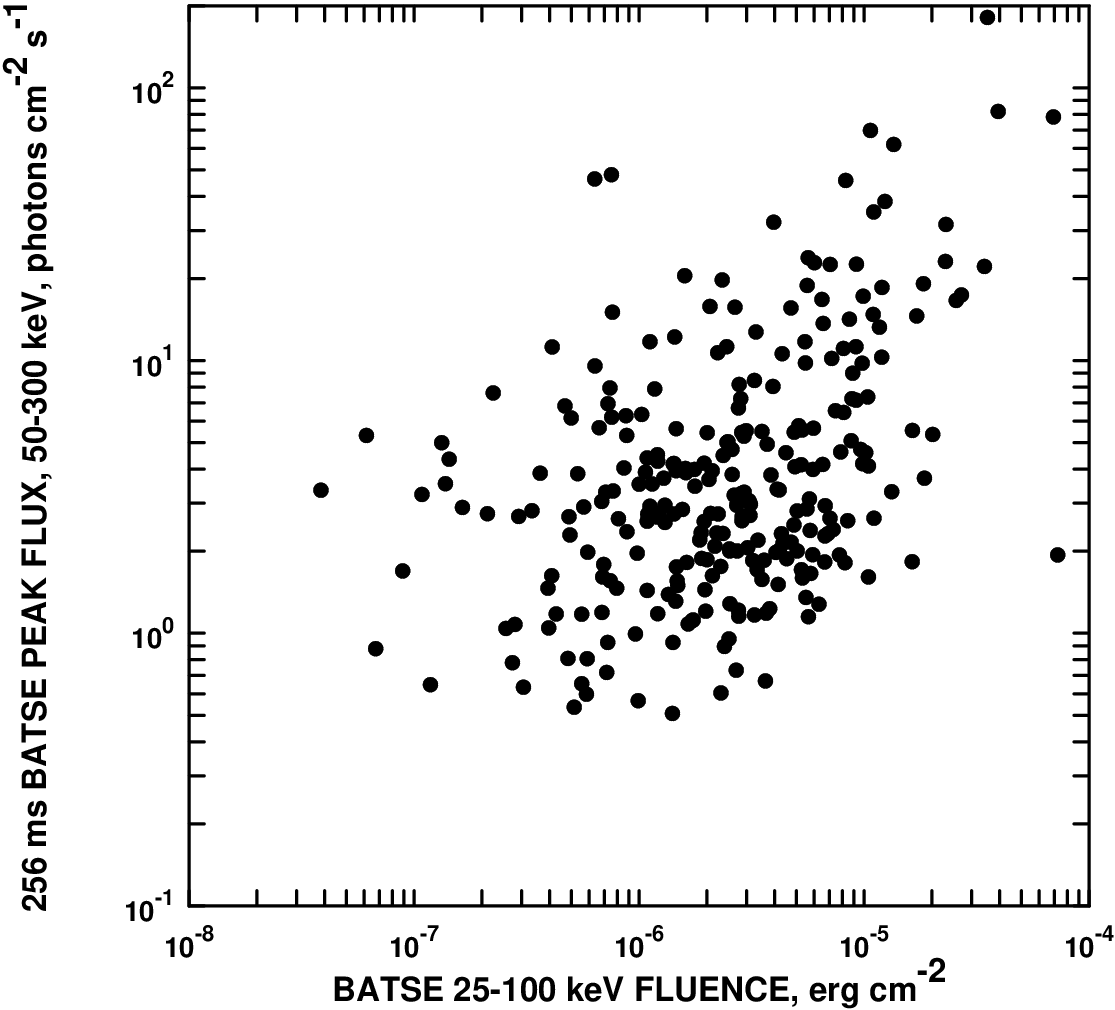}
\caption{Peak fluxes (measured over 256 ms, 50-300 keV) and 25-100 keV fluences of
272 of the bursts in this catalog.  No entries are given in the 5B catalog for 71
of the bursts.}
\end{figure}

\clearpage

\begin{deluxetable}{cccccccccccccc}

\rotate
\tabletypesize{\scriptsize}
\tablewidth{0pt}
\tablecaption{\it IPN annuli}
\tablehead{
\colhead{Date}&\colhead{UT}&\colhead{N$_B$}&\colhead{$\alpha_{2000, B}$}&
\colhead{$\delta_{2000, B}$}&\colhead{$\sigma_{stat, B}$}&\colhead{$\alpha_{2000, IPN1}$}&
\colhead{$\delta_{2000, IPN1}$}&\colhead{R$_{IPN1}$}&\colhead{$\delta R_{IPN1}$}&
\colhead{$\alpha_{2000, IPN2}$}&
\colhead{$\delta_{2000, IPN2}$}&\colhead{R$_{IPN2}$}&\colhead{$\delta R_{IPN2}$}
}
\startdata

960831 & 10:28:30 & 5591 & 242.76 & 22.93 & 1.01 & 166.481 & 36.669 & 66.664 & 0.044 & -- & -- & -- & -- \\
960906 & 22:39:48 & 5593 & 65.10 & -41.47 & 0.80 & 347.768 & -35.838 & 59.726 & 0.036 & -- & -- & -- & -- \\
960907 & 06:50:11 & 5594 & 143.93 & 49.87 & 3.72 & 167.836 & 35.799 & 17.877 & 0.293 & -- & -- & -- & -- \\
960912 & 13:57:27 & 5601 & 95.14 & 76.86 & 0.64 & 168.864 & 35.168 & 52.499 & 0.021 & -- & -- & -- & -- \\
960913 & 23:05:18 & 5604 & 182.58 & -23.45 & 0.92 & 169.129 & 35.010 & 57.377 & 0.066 & -- & -- & -- & -- \\
960917 & 22:22:32 & 5606 & 313.87 & -33.89 & 0.70 & 349.881 & -34.568 & 28.471 & 0.094 & -- & -- & -- & -- \\
960921 & 15:03:48 & 5609 & 345.27 & -28.55 & 0.21 & 350.571 & -34.178 & 5.417 & 0.030 & -- & -- & -- & -- \\
960924 & 11:41:50 & 5614 & 37.28 & 2.70 & 1.53 & 351.097 & -33.892 & 54.760 & 0.003 & -- & -- & -- & -- \\
\enddata
\end{deluxetable}
\clearpage
\noindent
a Degenerate localization \\
b Localized by the \it Rossi X-Ray Timing Explorer \rm All Sky Monitor 
(RXTE - Smith et al. 1999a) \\
b Localized by the \it Advanced Satellite for Cosmology and Astrophysics \rm
(ASCA - Murakami et al. 1997b) \\
d Localized by \it BeppoSAX \rm (Galama et al. 1997) \\
e Localized by \it BeppoSAX \rm (Piro et al. 1997a) \\
f Localized by the RXTE Proportional Counter
Array (Marshall et al. 1997) and by ASCA (Murakami et al. 1997a) \\
g Localized by \it BeppoSAX \rm (Antonelli et al. 1997) \\
h Localized by \it BeppoSAX \rm (Piro et al. 1997b) \\
i Localized by \it BeppoSAX \rm (in't Zand et al. 1998a) \\
j Localized by \it BeppoSAX \rm (Celidonio et al. 1998) \\
k Localized by \it BeppoSAX \rm (in't Zand et al. 1998b) \\
l Localized by \it BeppoSAX \rm (Nicastro et al. 1998) \\
m Localized by \it BeppoSAX \rm (Galama et al. 1998) \\
n Localized by \it BeppoSAX \rm (Heise et al. 1999) \\
o Localized by the RXTE All Sky Monitor (Smith
et al. 1999b) \\
p Localized by the RXTE Proportional Counter
Array (Takeshima and Marshall 1999) \\
q Localized by \it BeppoSAX \rm (Dadina et al. 1999) \\
r Localized by \it BeppoSAX \rm (Piro et al. 1999a) \\
s Localized by \it BeppoSAX \rm (Piro et al. 1999b) \\
t Localized by \it BeppoSAX \rm (in't Zand et al. 2000) \\
u Localized by BeppoSAX (Smith et al. 1999) \\
v Localized by the RXTE Proportional Counter 
Array (Takeshima et al. 1999) and by the \it Chandra X-Ray Observatory \rm (Piro et al. 1999c) \\

\clearpage
\begin{deluxetable}{cccc}
\tablecaption{\it IPN error boxes}
\tablewidth{0pt}
\tablehead{
\colhead{Date}&\colhead{N$_B$}&\colhead{$\alpha_{2000}$}&
\colhead{$\delta_{2000}$}
}
\startdata

970912 & 6380 &    
 11   $^{\rm h}$   57 $^{\rm m}$ 49.30 $^{\rm s}$ &   -22 $^{\rm o}$ 39 \arcmin  16.50 \arcsec   \\
 & &  11   $^{\rm h}$   53 $^{\rm m}$ 29.31 $^{\rm s}$ &   -23 $^{\rm o}$  6 \arcmin  46.07 \arcsec   \\
 & &  12   $^{\rm h}$    2 $^{\rm m}$  2.79 $^{\rm s}$ &   -22 $^{\rm o}$ 18 \arcmin  32.45 \arcsec   \\
 & &  11   $^{\rm h}$   53 $^{\rm m}$ 36.51 $^{\rm s}$ &   -22 $^{\rm o}$ 58 \arcmin   8.65 \arcsec   \\
 & &  12   $^{\rm h}$    2 $^{\rm m}$ 10.11 $^{\rm s}$ &   -22 $^{\rm o}$  9 \arcmin  37.02 \arcsec   \\

970919 & 6389 &   
  9   $^{\rm h}$   14 $^{\rm m}$ 51.83 $^{\rm s}$ &   -6 $^{\rm o}$ 29 \arcmin  50.75 \arcsec   \\
  & &  9   $^{\rm h}$   15 $^{\rm m}$ 12.79 $^{\rm s}$ &   -6 $^{\rm o}$ 37 \arcmin  34.39 \arcsec   \\
 & &   9   $^{\rm h}$   14 $^{\rm m}$ 25.88 $^{\rm s}$ &   -6 $^{\rm o}$ 25 \arcmin   0.55 \arcsec   \\
\enddata
\end{deluxetable}

\end{document}